\documentclass[aps,prd,showpacs,groupedaddress]{revtex4}
\usepackage{graphicx}% Include figure files
\usepackage{dcolumn}% Align table columns on decimal point
\usepackage{bm}% bold math
\usepackage{natbib}% bold math
\usepackage{lineno}

\begin{document}

%\preprint{}


\title{Measurement of the Resonance Parameters of the $\chi_{1}(1^3P_1)$
and $\chi_{2}(1^3P_2)$ States of Charmonium formed in Antiproton-Proton
Annihilations}


\author{M.~Andreotti,$^2$
S.~Bagnasco,$^{3,6}$
W.~Baldini,$^2$  
D.~Bettoni,$^2$ 
G.~Borreani,$^6$  
A.~Buzzo,$^3$ 
R.~Calabrese,$^2$  
R.~Cester,$^6$
G.~Cibinetto,$^2$ 
P.~Dalpiaz,$^2$  
G.~Garzoglio,$^1$
K.~E.~Gollwitzer,$^1$   
M.~Graham,$^7$
M.~Hu,$^1$
D.~Joffe,$^5$
J.~Kasper,$^5$  
G.~Lasio,$^4$    
M.~Lo Vetere,$^3$
E.~Luppi,$^2$    
M.~Macr\`\i,$^3$  
M.~Mandelkern,$^4$  
F.~Marchetto,$^6$  
M.~Marinelli,$^3$ 
E.~Menichetti,$^6$ 
Z.~Metreveli,$^5$
R.~Mussa,$^{2,6}$  
M.~Negrini,$^2$
M.~M.~Obertino,$^{6,7}$
M.~Pallavicini,$^3$  
N.~Pastrone,$^6$  
C.~Patrignani,$^3$ 
S.~Pordes,$^1$
E.~Robutti,$^3$
W.~Roethel,$^{4,5}$
J.~Rosen,$^5$  
P.~Rumerio,$^5$ 
R.W.~Rusack,$^7$
A.~Santroni,$^3$ 
J.~Schultz,$^4$ 
S.H.~Seo,$^7$
K.~K.~Seth,$^5$ 
G.~Stancari,$^{1,2}$ 
M.~Stancari,$^{2,4}$
A.~Tomaradze,$^5$  
I.~Uman,$^5$
T.~Vidnovic,$^7$
S.~Werkema,$^1$
P.~Zweber$^5$ \\
\vspace*{0.2cm}
(Fermilab E835 Collaboration)  \\
 }

\affiliation{\vspace*{0.2cm}
$^1$Fermi National Accelerator Laboratory, Batavia, Illinois 60510 \\
$^2$Istituto Nazionale di Fisica Nucleare and University of Ferrara, 
44100 Ferrara, Italy \\
$^3$Istituto Nazionale di Fisica Nucleare and University of Genova, 
16146 Genova, Italy \\
$^4$University of California at Irvine, California 92697 \\
$^5$Northwestern University, Evanston, Illinois, 60208 \\
$^6$Istituto Nazionale di Fisica Nucleare and University of Torino,
10125, Torino, Italy\\
$^7$University of Minnesota, Minneapolis,Minnesota 55455 \\}
\begin{abstract}
%\date{\today}
%\begin{abstract}

We have studied the $^3P_J$ ($\chi_c)$ states of charmonium in formation
by antiproton-proton annihilations in experiment E835 at the Fermilab
Antiproton Source. We report new measurements of the mass, width, and
$B(\chi_{cJ} \rightarrow \bar{p} p) \times \Gamma(\chi_{cJ} \rightarrow
J/\psi + ~anything)$ for the $\chi_{c1}$ and $\chi_{c2}$ by means of the
inclusive reaction $\bar{p}p~\rightarrow~\chi_{cJ} ~\rightarrow~
J/\psi~+~anything \rightarrow (e^{+}e^{-}) +anything $.  Using the
subsample of events where $\chi_{cJ} \rightarrow \gamma + J/\psi
\rightarrow \gamma + (e^{+}e^{-}) $ is fully reconstructed, we derive
$B(\chi_{cJ} \rightarrow \bar p p)\times \Gamma(\chi_{cJ} \rightarrow
J/\psi + \gamma) $. We summarize the results of the E760 (updated) and
E835 measurements of mass, width and $ B(\chi_{cJ} \rightarrow \bar{p}p)
\Gamma(\chi_{cJ} \rightarrow J/\psi+\gamma)$ (J=0,1,2) and discuss the
significance of these measurements.

\end{abstract} 


%\keywords{}

\maketitle

% References should be done using the  \cite, \ref, and \label commands

%Line number
%\begin{linenumbers}

\section{Introduction}

Since the discovery of charmonium, it has been clear that the properties
of the $^3P_J $ ($\chi_{cJ}$) states are key elements in the understanding
of the role and limitations of perturbative Quantum Chromodynamics (pQCD)
in this energy regime.  The existence of a triplet of P states, split by
spin-orbit and tensor force terms, allows us to probe the spin structure
of QCD forces. 
  
The production and decay mechanisms of the $\chi_{cJ}$ states are still
actively being studied at low energy $e^{+}e^{-}$ storage rings, at high
energy colliders and in fixed target 
experiments\cite{Brambilla}. The most precise
determinations of mass and width come, however, from our study of
charmonium spectroscopy by formation of $\bar{c} c$ states in $\bar{p}p$
annihilation at the Fermilab Antiproton Source (experiments E760 and
E835).  The E760 collaboration measured the resonance parameters of the
$\chi_{c1}$ and $\chi_{c2}$\cite{chi12} and more recently we reported
measurements of the $\chi_{c0}$ \cite{chi0}. In this paper we present the
results of new measurements of the $\chi_{c1}$ parameters made, with
greatly improved statistics, by E835.  The $\chi_{c2}$ parameters were
also remeasured with statistics comparable to those of experiment E760.
Our E760 results have been updated to account for revised values of 
reference parameters and are quoted below.

\section{Experimental technique}

We briefly review the technique used in this experiment. A localized
source (0.5$\times$0.5$\times$0.6 cm$^3$) of $\bar{p}p$ interactions at
instantaneous luminosities up to $ 5 \times 10^{31} $cm$^{-2} $s$^{-1}$
was obtained by intersecting the beam of stochastically cooled antiprotons
circulating in the Accumulator, with a jet of clusterized hydrogen
molecules ($\rho_{max} = 3.0 \times 10^{14} $atoms$/$cm$^{3})$. The
momentum of the antiproton beam was changed in small steps allowing a fine
scan of narrow resonances. The parameters of a $\bar{c}c$ resonance (R),
mass, width and $B_{R\rightarrow\bar{p}p}\times \Gamma_{R\rightarrow
final~state}$, were then determined from the excitation curve obtained by
measuring the cross section at each value of the antiproton-proton
center-of-mass energy ($\sqrt s$).  With this technique, the systematic
uncertainties in the mass and width measurements are greatly reduced since
they depend only on the knowledge of the center-of-mass energy.

We determine the center-of-mass energy distribution by measuring the
beam-revolution-frequency spectrum and the orbit length, as described in
detail in reference \cite{psipaper}. We calibrate the central orbit length
$L_0$ using the recent high-precision measurement of the $\psi'$ mass
by the KEDR experiment, $3686.111 \pm 0.025 \pm 0.009$ MeV/c$^2$ \cite{kedr},
which gives an uncertainty of $\pm 0.17$ mm out of 474.046 m.  
$\Delta L$, the correction to $L_0$ due to deviations from the central
orbit, is determined using 48 horizontal beam-position monitors (BPMs)
\cite{bigpaper} \cite{psipaper}. For scan I at the $\chi_{c1}$, the
uncertainty in $\Delta L$ was estimated as 1 mm (rms) [110 keV]
\cite{psipaper}. The BPM system was subsequently improved and we estimate
the uncertainty for the subsequent scans as 0.64 mm (rms) [70 keV at the
$\chi_{c1}$, 75 keV at the $\chi_{c2}$]. The center-of-mass energy spread,
$\sigma _{\sqrt s}$, was approximately 200 keV at the $\chi_c$ formation
energies.

The cross section for formation of the $\chi_c$
states is less than 10$^{-5}$ of the inelastic $\bar{p}p$ hadronic cross
section.  Even so, a clean signal was extracted by selecting
electromagnetic final states as tags of charmonium formation. The
$\chi_{cJ}$ were studied in the inclusive reaction:

\begin{equation}
\bar{p}p~\rightarrow~\chi_{cJ}~\rightarrow~ J/\psi + anything
\rightarrow~ (e^{+}e^{-}) + anything \, .
\protect{\label{eq:reac}}
\end{equation}

\begin{figure}[hbtp]
\vspace*{+0.5cm}
\begin{center}
\rotatebox{270}{\includegraphics[scale=0.35 ]{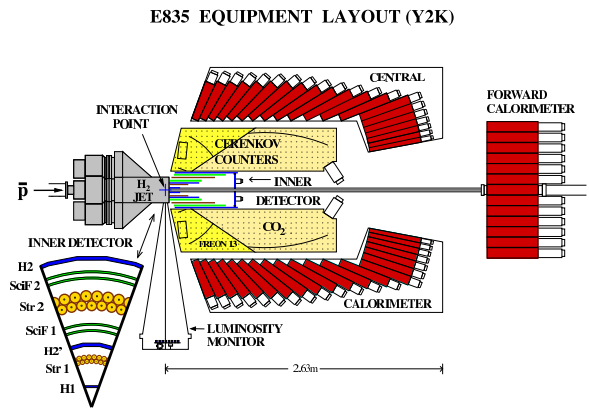}}
\caption{The E835 detector, side view} 
\label{fig:det}
\end{center}
\end{figure}

The non-magnetic spectrometer (Fig.\ref{fig:det}) was optimized for the
detection of photons and electrons, and is described in detail in
reference \cite{bigpaper}. The apparatus had full acceptance in azimuth
($\phi$), with a cylindrical central system and a planar forward system.
The detector elements used for the trigger and for the offline selection 
of
events from reaction (1) were (a) three hodoscopes, $H1$, $H2'$ and $H2$,
azimuthally segmented in 8, 24 and 32 counters respectively, (b) a
threshold gas \v{C}erenkov counter for identifying $e^\pm$, divided in two
volumes in polar angle; each volume was segmented azimuthally in 8 sectors
aligned with the counters of the $H1$ hodoscope, and (c) two lead-glass
calorimeters for measuring the energy and direction of photons and
electrons: a cylindrical one (CCAL) with 1280 counters, covering the polar
angles $ 11^\circ < \theta < 70^\circ $ and a planar one (FCAL)  covering
the polar angles $ 3^\circ < \theta < 12^\circ $. All counters were
equipped with time and pulse-height measurement capability. The luminosity
was measured at each data point with a statistical precision of $0.1\%$
and systematic uncertainty of $\pm 2.5 \% $, by counting recoil protons
from elastic $\bar{p}p$ scattering in three solid state detectors located
at $87.5^\circ$ to the beam direction.

The hardware trigger was designed to select events with a $J/\psi
\rightarrow e^+e^-$ decay in the central detector \cite{charge}. It
required two charged tracks, each defined by a coincidence between two
hodoscope counters ($H1 \times H2$) aligned in azimuth, with at least one 
of
the two particles tagged as an electron by a signal in the corresponding
\v{C}erenkov cell.  In addition, two large energy deposits (clusters)
separated by more than 90$^\circ$ in azimuth and with an invariant mass
greater than 60$\%$ of the center of mass energy, were required in the
CCAL. The efficiency of this trigger was measured to be $ 0.90\pm 0.02$
from a clean sample of $ \bar{p} p \rightarrow \psi' \rightarrow e^+e^-$
events, taken with relaxed trigger conditions. Online, a filtering program
certified as electron candidates CCAL energy clusters aligned with tracks
formed by the hodoscopes and \v{C}erenkov elements.

\section{Data analysis}

The data presented here were collected by experiment E835, in three scans
performed at the $\chi_{c1} $ in August 1997 (scan I), February 2000 (scan
II), and July 2000 (scan III), and one scan performed at the $\chi_{c2} $
in February 2000. The center of mass energy, $\sqrt{s}$, width of the
$\sqrt s$ distribution, $FWHM_{\sqrt{s}}$, and integrated luminosity,
${\it Ldt}$, for each $run$ are given in Table \ref{tab:dati2}.

\begin{figure}[htb] \vspace*{+0.5 cm} \begin{center} \begin{tabular}{c}
\includegraphics[height=6.1cm,width=7.3cm]{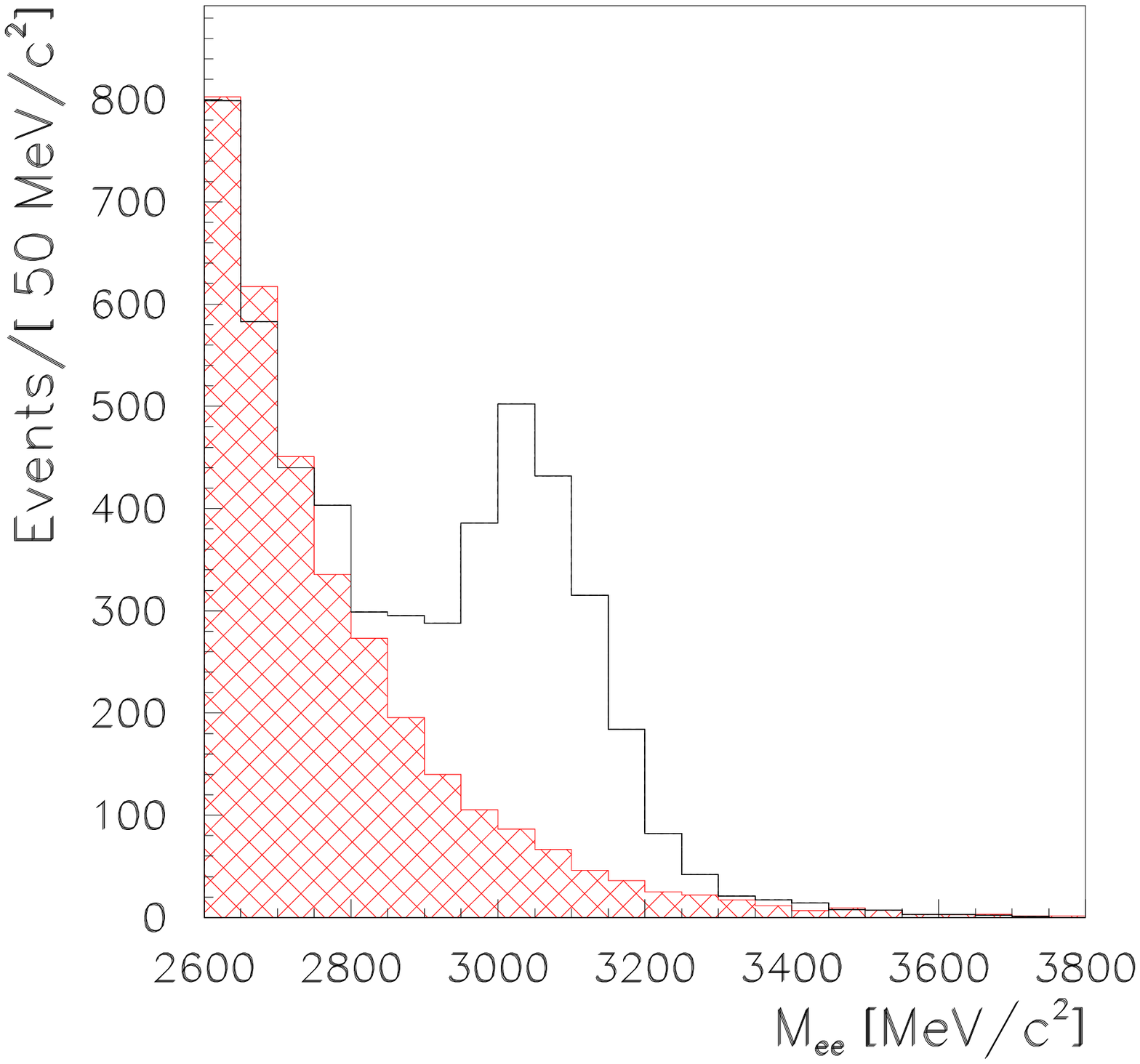} \\
\includegraphics[height=6.1cm,width=7.3cm]{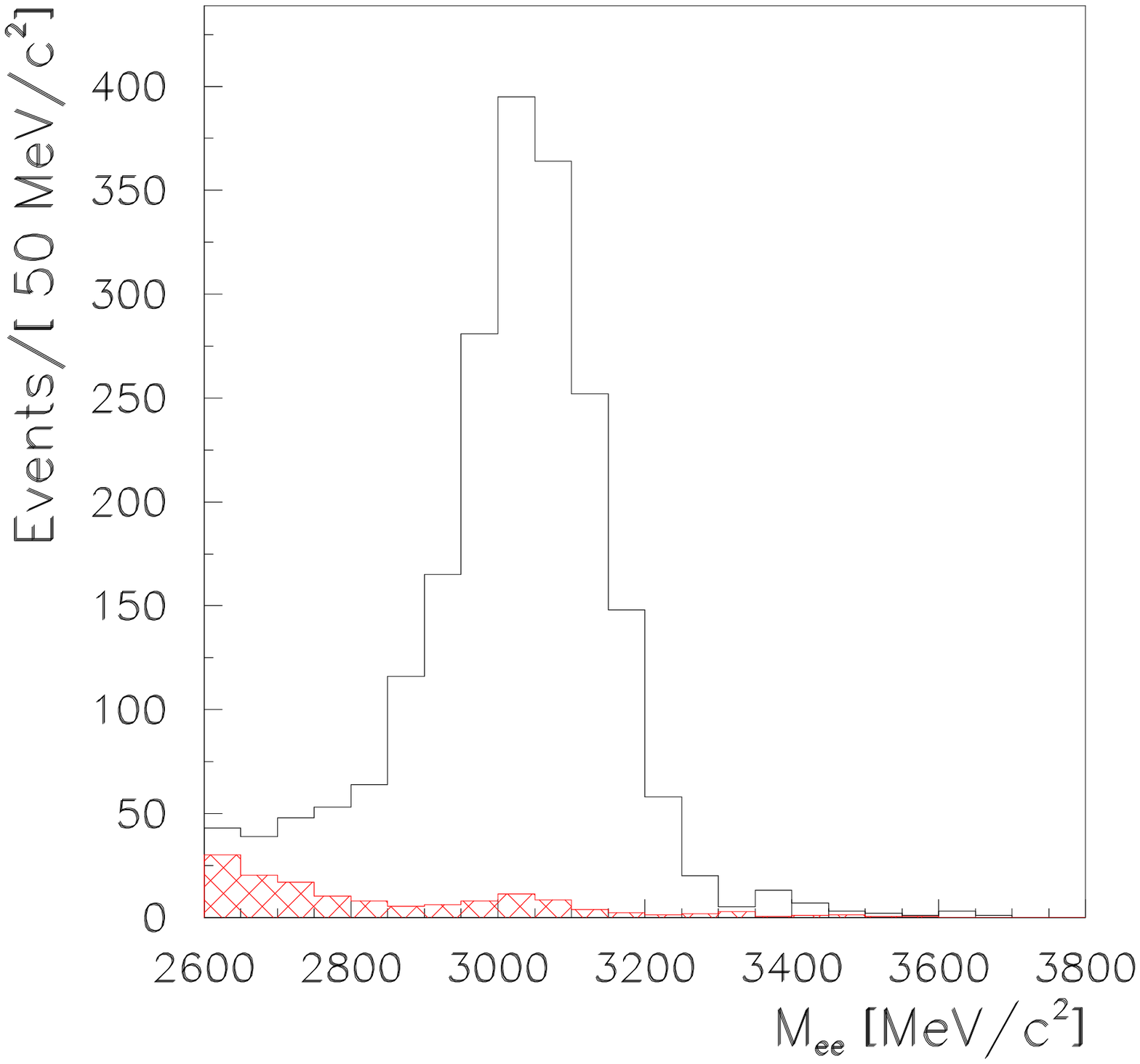} \\
\end{tabular} 

\caption{Reconstructed e$^+$e$^-$ invariant-mass distribution for events
in the $\chi_{c1}$ resonance region(clear), and for events
off-resonance(shaded): (a) all events with both electron candidates within
the \v{C}erenkov fiducial volume, (b) events remaining after applying the
electron-weight cut.}

\label{fig:mee} \end{center}
\end{figure}

Our data analysis methods are described in detail in reference
\cite{bigpaper}. The offline selection of $\chi_{c1}$ and $\chi_{c2}$
events compatible with reaction (1) is done in three steps; the first two
are illustrated in Fig. \ref{fig:mee} which shows, at each step, the
invariant mass distribution of the $e^+~e^-$ candidates for on-resonance
data and (shaded) for data taken off-resonance, normalized to the
integrated luminosity of the data taken in the resonance region.  In the
first step (Fig. \ref{fig:mee}a), all events with two electron 
(i.e. electron-positron)
candidates within the \v{C}erenkov fiducial region ($15^\circ<\theta <
60^\circ $) and an invariant mass ($M_{ee}$) above 2600 MeV/c$^2$ are
selected.  A clear enhancement is seen in the on-resonance data at the
mass of the $J/\psi$. In the next step, the electron candidates are
identified by using an ``electron weight" parameter, which is a likelihood
ratio for the electron hypothesis versus the background hypothesis. It
uses the pulse heights in the three hodoscopes ($H1$,$H2'$,$H2$) and
\v{C}erenkov counter, and the transverse energy distribution of the CCAL
clusters, and distinguishes single electron tracks from background
(predominantly $e^{+}e^{-}$ pairs from photon conversions in the 0.18 mm
thick steel beam-pipe and $\pi^0\rightarrow e^+ e^-\gamma$ decays). The
resulting $e^{+}e^{-}$ invariant mass distribution is shown in Fig.  
\ref{fig:mee}b. The slight enhancement in the background level at the
$J/\psi$ mass peak comes from the continuum of $\bar p p \rightarrow \pi^0
J/\psi $ events.

We select $J/\psi X$ events by applying a 1C kinematical fit to the
reaction: $\bar{p}p\rightarrow\chi_{cJ}\rightarrow
J/\psi+anything\rightarrow(e^+e^-)+anything $, accepting events with
$\chi^2$ probability greater than $10^{-2} $ and with $M_{ee}> 2800$
MeV/c$^2$. The number $N(J/\psi X)$ of events selected for each run is given in
Table \ref{tab:dati2}. The efficiency for event selection is determined
from a sample of events collected in a run ({labeled \it efficiency} in
Table \ref{tab:dati2}) taken near the $\chi_{c1}$ resonance peak energy,
and is 0.865 $\pm$ 0.015. The geometrical acceptances are calculated by
Monte Carlo simulation, fixing the parameters of the angular distributions
of reaction (1) to values measured previously \cite{angdis}. The
acceptances are $ 0.610 \pm 0.006 $ for the $\chi_{c1} $ and $ 0.617 \pm
0.006 $ for the $\chi_{c2} $, where the errors include uncertainties in
the acceptance-volume boundaries and angular-distribution parameters.  
After including trigger and selection efficiencies, we obtain the overall
efficiencies ($\epsilon$) given in Table \ref{tab:chi1res}.

We select $J/\psi\gamma$ events from the inclusive sample by requiring one
additional on-time cluster within the fiducial volumes of CCAL
(12$^\circ<\theta<$68$^\circ$) or FCAL (3$^\circ<\theta<$10$^\circ$). [For
scan I at the $\chi_{c1}$ we did not use FCAL.] A 5C kinematic fit to the
reaction: 
\begin{equation} \bar{p} p\rightarrow
J/\psi+\gamma \rightarrow(e^+e^-)+ \gamma \protect{\label{eq:rad}}
\end{equation} 
is applied and events with $\chi^2$ probability less than
$10^{-3} $ are rejected. The number $N(J/\psi \gamma)$ of events selected
for each run is given in Table \ref{tab:dati2}. The geometrical
acceptances are $0.498 \pm 0.01$ for the $\chi_{c1}$ [$0.456 \pm 0.01$ for
scan I] and $0.519\pm 0.01$ for the $\chi_{c2}$. The overall efficiencies
($\epsilon$) are given in Table \ref{tab:chi1res}.

\begin {table}[htb]
\begin{center}
\begin {tabular}{|cccccc|}
\hline
 & $\sqrt{s} $ & $FWHM_{\sqrt{s}} $ & {\it Ldt} 
&   N($J/\psi$ X)  &  N$(J/\psi \gamma) $ \\
 & [MeV] & [keV] & [nb$^{-1}$] & & \\
\hline\hline
 &   3513.00 & 713 & 301.3  & 24 & 14\\
$\chi_{1}$ &  3511.44  & 740  & 315.5  & 110 & 77  \\
 scan I &   3511.05 & 723 & 319.4 & 178 & 120\\
 Aug. 97 &   3510.75 & 682 & 318.8 & 266 & 175\\
 &   3510.36 & 656 & 315.0 & 217 & 151\\
 &   3509.93 & 592 & 317.1 & 101 & 66\\
 &   3508.59 & 545 & 376.2 & 20 & 11\\
 &   3494.43 & 788 & 502.8 & 10 & 2\\
 &   3524.64 & 717 & 3716.9 & 57 & 14\\
 &   3525.16 & 661 & 2903.0 & 53 & 20\\
\hline \hline
       &    3511.79  &  635 & 184.9 &   36     &   27	   \\
       &    3511.39  &  566 & 200.9 &   84     &   62	   \\
$\chi_{1}$ &    3511.03  &  562 & 190.8 &  139     &  101	   \\
 scan II      &    3510.56  &  550 & 199.5 &  182     &  129	   \\
 Feb. 00      &    3510.15  &  512 & 235.3 &  122     &   83	   \\
       &    3509.74  &  457 & 319.0 &   67     &   48	   \\
\hline \hline
 &   3511.69  &  727 & 452.1  &   77     &   47       \\
$\chi_{1}$ &   3510.69  &  675 & 417.6  &   338    &  241       \\
 scan III &   3511.17  &  746 & 441.0  &   214    &  147       \\
July 00 &   3510.21  &  604 & 493.4  &   333    &  245        \\
 &   3509.69  &  472 & 750.2  &   186    &  140        \\
\hline \hline
efficiency  &   3510.62  &  721 & 1874.2 &  1422    & 1049        \\
\hline \hline
 &    3558.80  &  533 & 144.4 &   20     &   19	   \\
 &    3557.31  &  519 & 205.5 &   86     &   70	   \\
$\chi_{2}$ &    3555.82  &  481 & 267.0 &  248     &  191	   \\
 Feb. 00 &    3554.29  &  439 & 225.4 &   54     &   49	   \\
 &    3535.10  &  444 & 211.0 &    5     &    1	   \\
\hline \hline
 &   3469.90   & 802 & 2512.6  &     20   &  2	 \\
background &   3525.17  &  708 & 3709.6  &     44   & 11	   \\
00 &   3523.33  &  920 & 3058.6  &     49  & 17	 \\
 &   3524.79  &  701 & 2033.0  &     33  & 13	 \\
\hline
\end {tabular}
\vspace{+0.5cm}

\caption{Center of mass energy ($\sqrt{s}$) and width of $\sqrt{s}$
distribution ($FWHM_{\sqrt{s}}$), integrated luminosity {\it Ldt}, and
number of events selected (N) for the inclusive ($J/\psi X$) and exclusive
($ J/\psi \gamma $) decay channels for each data run used in this
analysis.}

\protect{\label{tab:dati2}}
\end{center}
\end{table}

Scan I, performed in 1997, includes three background points distant from
the $\chi_{c1}$. The entries labeled {\it background} in Table
\ref{tab:dati2} refer to data taken in 2000 that are far from the
$\chi_{c}$ resonances. The background point at $\sqrt{s}$ = 3469.9 MeV
is used only for the $\chi_{c1}$ analysis while the other three points
are used for both $\chi_{c1} $ and $\chi_{c2} $.\\

\begin {table*}[htb]
\begin {tabular}{|ccc|}
\hline\hline

&   &   \\
& {\large $\chi_{c1}$} Scan I  & \\
 &   &   \\
\hline
 & & \\
 & $\bar p p \rightarrow \chi_{c1} \rightarrow J/\psi + X $~~~ &
 $\bar p p \rightarrow \chi_{c1} \rightarrow J/\psi + \gamma$ \\
 & $J/\psi \rightarrow e^+ e^-$ & $J/\psi \rightarrow e^+ e^-$ \\
 &   &  \\

$M_{\chi_{c1}}$ [MeV/c$^2$] & $3510.749\pm0.122$ & 
$3510.749\pm0.113$ \\ 
$\Gamma_{\chi_{c1}}$ [MeV] & $0.89\pm 0.09$ 
& $0.89 \pm 0.09 $ \\
B$_{in}\times \Gamma_{out}$[eV]  &$1.25\pm0.06$ & 
$1.20\pm0.06$   \\
$\sigma_b$[pb] & $14.5\pm 1.6$  & $3.4\pm0.9$ \\
$\epsilon$ & $0.459\pm 0.011$ & $0.328\pm0.010$ \\ 
$\chi^2$/D.F.   &7.2/6   &  11.5/14 \\
\hline\hline

 &   &   \\
& {\large $\chi_{c1}$} Scan II  & \\
 &   &   \\
\hline
 & & \\
 & $\bar p p \rightarrow \chi_{c1} \rightarrow J/\psi + X $~~~ &
 $\bar p p \rightarrow \chi_{c1} \rightarrow J/\psi + \gamma$ \\
 & $J/\psi \rightarrow e^+ e^-$ & $J/\psi \rightarrow e^+ e^-$ \\
 &   &  \\
$M_{\chi_{c1}}$ [MeV/c$^2$]   & 
$3510.783\pm 0.075$   &  $3510.784\pm0.075$ \\ 
$\Gamma_{\chi_{c1}}$ [MeV]  & 
$0.87\pm 0.08$   &    $0.88\pm 0.09 $   \\
B$_{in}\times \Gamma_{out}$[eV]  & 
$1.33 \pm 0.06 $ & $1.22\pm0.06$ \\
$\sigma_b$[pb] & $10.4 \pm 1.1$ &  $2.2 \pm 0.6$ \\
$\epsilon$ & $0.475 \pm 0.015$ & $0.382 \pm 0.014$ \\ 
$\chi^2$/D.F.   & 5.6/6  &   12.3/14  \\
\hline\hline

 &   &   \\
& {\large $\chi_{c1}$} Scan III  & \\
 &   &   \\
\hline
 & & \\
 & $\bar p p \rightarrow \chi_{c1} \rightarrow J/\psi + X $~~~ &
 $\bar p p \rightarrow \chi_{c1} \rightarrow J/\psi + \gamma$ \\
 & $J/\psi \rightarrow e^+ e^-$ & $J/\psi \rightarrow e^+ e^-$ \\
 &   &  \\
$M_{\chi_{c1}}$ [MeV/c$^2$] & $3510.643\pm 0.074$ &
$3510.641\pm 0.074$ \\
$\Gamma_{\chi_{c1}}$ [MeV]  & 
$0.87\pm0.06 $   &  0.$88 \pm0.07$ \\
B$_{in}\times \Gamma_{out}$[eV]  & 
$1.25\pm0.04$ & $1.14\pm0.04$  \\
$\sigma_b$[pb] &  $10.5\pm1.1$ &  $2.3\pm0.6$\\
$\epsilon$ & $0.475\pm 0.015 $ & $0.382\pm0.014$ \\
$\chi^2$/D.F.       &6.4/5    &17.3/12 \\
\hline\hline

&   &   \\
& {\large $\chi_{c2}$}  & \\
 &   &   \\
\hline
 & & \\
 & $\bar p p \rightarrow \chi_{c2} \rightarrow J/\psi + X $~~~ &
 $\bar p p \rightarrow \chi_{c2} \rightarrow J/\psi + \gamma$ \\
 & $J/\psi \rightarrow e^+ e^-$ & $J/\psi \rightarrow e^+ e^-$ \\
 &   &  \\
$M_{\chi_{c2}}$ [MeV/c$^2$]   &$3556.173\pm0.123$ &$3556.168\pm0.114$ \\ 
$\Gamma_{\chi_{c2}}$ [MeV]  & $1.92 \pm0.19$ & $1.95\pm0.19 $\\
B$_{in}\times \Gamma_{out}$[eV]  & $1.60\pm0.09$& $1.59\pm0.10$ \\
$\sigma_b$[pb] & $12.0\pm1.3$& $2.6\pm0.7 $  \\
$\epsilon $ & $0.481 \pm 0.017 $ & $ 0.398\pm 0.014$ \\ 
$\chi^2$/D.F.   & 3.6/4 & 18.9/10    \\
\hline\hline

\end{tabular}
\vspace{+0.5cm}

\caption{Fit results for $\chi_{c1}$ and $\chi_{c2}$ parameters, listed
separately by scan. The $J/\psi X$ results are obtained by fitting the
inclusive sample. The $J/\psi\gamma$ results are obtained from a
joint fit to the $J/\psi \gamma$ sample and the remaining $J/\psi X$
events (see text). The errors include fitting uncertainties and random 
uncertainties in the beam-orbit length.}

\protect{\label{tab:chi1res}}
\end{table*}

The cross section:  $\sigma_{meas}(\sqrt{s_i}) = \frac{N_i}{{\it
L_i}\times\epsilon}$, measured at the $i$th point of a scan, is given by:
 \begin{equation} 
\sigma_{meas}(\sqrt{s_i}) = \sigma_b +
\int~[f_i(\sqrt{s_i}-\sqrt{s'})\times
~\sigma_{BW}^{rad}(\sqrt{s'})]~d\sqrt{s'} \,~~~~~~~~~~
\protect{\label{eq:sigma}} 
\end{equation} 
 where $\sigma_b $ is the background cross section, which we take to be
constant over each scan, $ f_i(\sqrt{s_i}-\sqrt{s'})$ is the normalized
$\sqrt s$ distribution at the {\it i}th point and $\sigma_{BW}^{rad} $
is the Breit-Wigner resonance cross section corrected for initial state
radiation (see appendix A). The Breit-Wigner cross section is:

\begin{equation}
\sigma_{BW}(\sqrt{s}) 
 = \frac{\pi(2J
+1)}{k^2} \times \frac{\Gamma_{\chi_{cJ}} \times B_{in} \times 
\Gamma_{out}}
{4(\sqrt{s}-M_{\chi_{cJ}})^2+\Gamma_{\chi_{cJ}}^2} ~~~~
\protect{\label{eq:BW}}
\end{equation}
where $k^2 = \frac{s-4m_{p}^2}{4}$, $m_p$ is the 
proton mass, $J$, $ M_{\chi_{cJ}}$ and 
$\Gamma_{\chi_{cJ}}$  are the spin, mass and 
width of the $\chi_{cJ}$ resonance,  
$B_{in} = B(\chi_{cJ}\rightarrow \bar{p}p)$ and 
$\Gamma_{out} = \Gamma(\chi_{cJ} \rightarrow J/\psi + ~anything~) \times 
B(J/\psi \rightarrow e^+ e^- $). 

A maximum likelihood fit to equation (\ref{eq:sigma}) is performed to find
the values of $M_{\chi_{cJ}}$, $\Gamma_{\chi_{cJ}}$, $\sigma_b $, and
$B_{in}\times \Gamma_{out}$. This last parameter effectively measures the
area of the Breit-Wigner since $B_{in}\times \Gamma_{out}$ can be
rewritten as $ (B_{in} B_{out})\times \Gamma_{\chi_{cJ}}$ and ($ B_{in}
B_{out}$) measures the cross section at the peak of the resonance.  The
errors in $M_{\chi_{cJ}}$ are the in-quadrature sums of uncertainties from
the maximum-likelihood fits and uncertainties in corrections to the
beam-orbit length, which is used in the determination of the
center-of-mass energy as described above. To determine the value of
$B_{in}\times \Gamma(\chi_{c1}\rightarrow J/\psi + \gamma) \times B(
J/\psi \rightarrow e^+e^-) $ we perform a joint maximum likelihood fit to
equation (\ref{eq:sigma}) of the two independent samples of events: 1.
$J/\psi \gamma$ fits, and 2. $J/\psi X$ events not fitting $J/\psi \gamma$
(see Table \ref{tab:dati2}), constraining $M_{\chi_{c1}}$ and
$\Gamma_{\chi_{c1}}$ to be the same for the two samples and allowing
$\sigma_b$ and $B_{in} \Gamma_{out}$ to be different. The results of the
fits are given in Table \ref{tab:chi1res}.

\begin{figure}[hbtp] 
\vspace*{+0.5 cm}
\begin{center}  
\begin{tabular}{c}
\includegraphics[height=6.1cm,width=7.3cm]{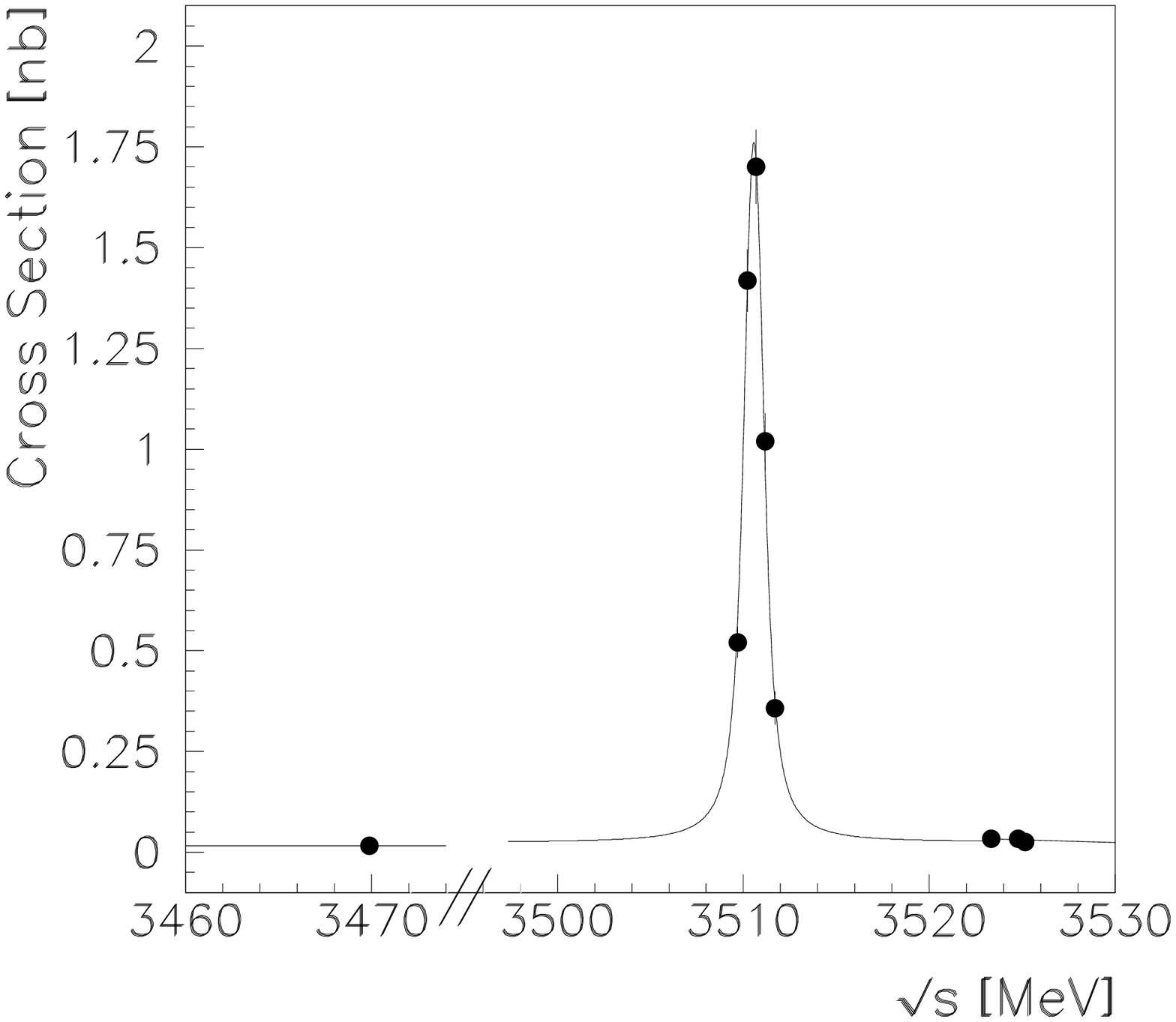} \\
\includegraphics[height=6.1cm,width=7.3cm]{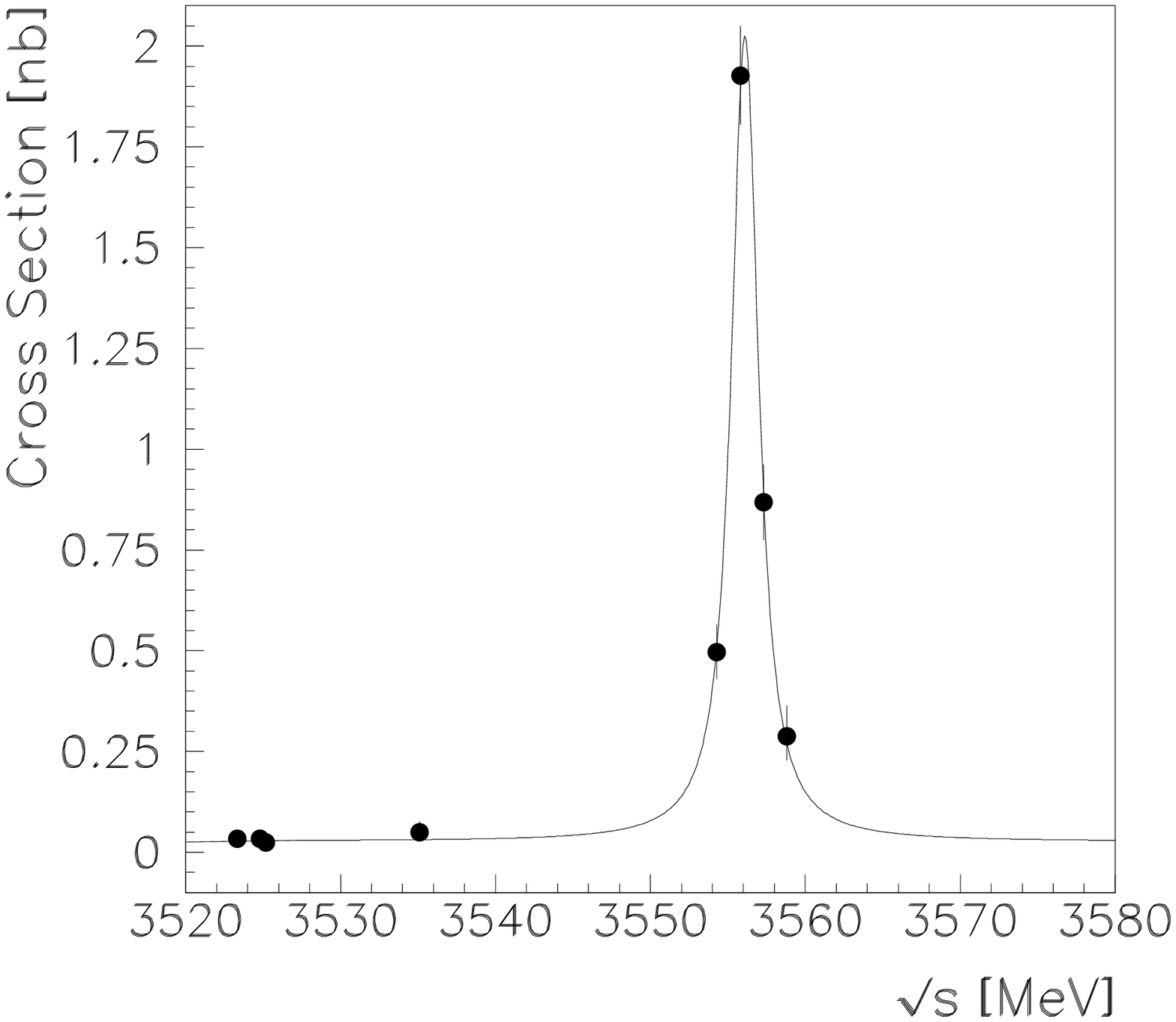} \\
\end{tabular}

\caption{(a)The measured cross sections and fitted excitation curve for
scan III of the $\chi_{c1}$ plus background points. (b) The measured cross
sections and excitation curve for the $\chi_{c2}$ scan plus background
points.}
 \label{fig:res}
 \end{center}
 \end{figure}

\begin{figure}[hbtp]
\begin{center}  
\includegraphics[height=8.1cm,width=9.0cm]{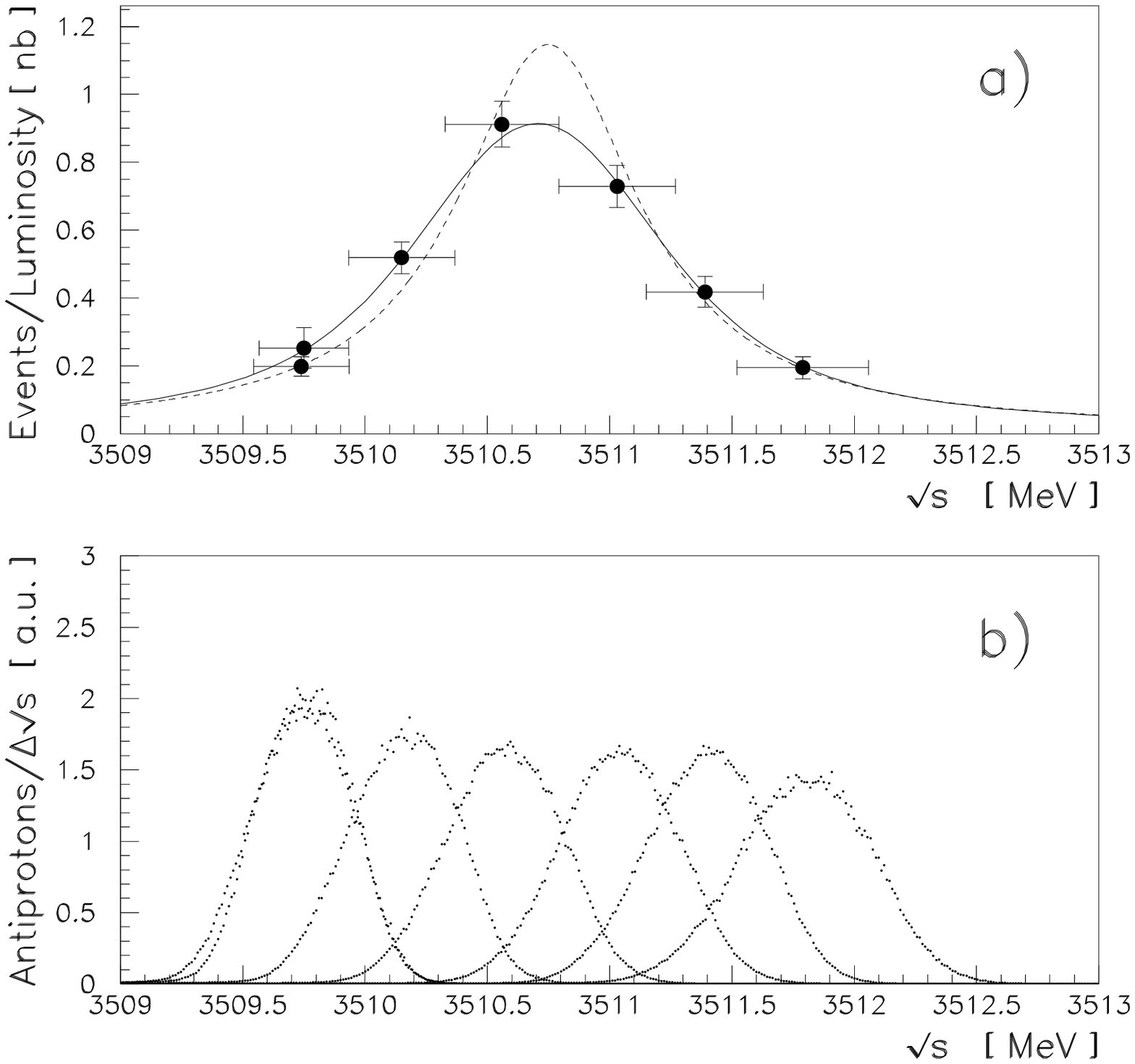} 

\caption{(a)The measured cross section for scan II at the $\chi_{c1}$. The
horizontal errors are the $FWHM_{\sqrt s}$ .  The solid curve is the fit
to Eq. \ref{eq:sigma}, and includes the spread in $\sqrt{s}$. The dashed
curve is the sum of $\sigma_b$ and the Breit-Wigner cross section
$\sigma_{BW}(\sqrt{s})$ given by Eq. (\ref{eq:BW}). (b)The $\sqrt s$
distribution for each point of the scan.}

\label{fig:giulio}
\end{center}
\end{figure}

In Fig. \ref{fig:res}a we plot, at each point of scan III at the
$\chi_{c1}$ plus background points, the measured cross section
($\sigma_{meas} $) superimposed on the excitation curve obtained from the
fitted parameters listed in Table \ref{tab:chi1res}, column 3. The same
graphical representation of the results of the $\chi_{c2}$ scan is given
in Fig. \ref{fig:res}b. To illustrate the effect of scanning a narrow
resonance with a beam of comparable width, we show in Fig.
\ref{fig:giulio}a a blow up of scan II at the $\chi_{c1}$, where the
horizontal errors are the $FWHM_{\sqrt s}$.  The solid curve is the fit to
Eq. \ref{eq:sigma}, and includes the spread in $\sqrt{s}$. The dashed
curve is the sum of $\sigma_b$ and the Breit-Wigner cross section
$\sigma_{BW}(\sqrt{s})$ given by Eq. (\ref{eq:BW}). The parameters for the
curves are given in Table \ref{tab:chi1res}, column 2.

\section{Results}

The results from the individual $\chi_{c1}$ scans are in good agreement
and therefore we take for each parameter the weighted (by the inverse
variance) average of the measurements. The resulting values of the
parameters for the two methods of determining them are given in Table
\ref{tab:chi12}, along with the corresponding $\chi_{c2}$ results. The
errors shown are (i) statistical, (ii) from uncertainties in {\it
auxiliary} variables, which were measured during data taking, and (iii)
from uncertainties in {\it external} parameters measured in other
experiments.  The last of these uncertainties may eventually be reduced.  
The auxiliary-variable error in $\Gamma_{\chi_{cJ}}$ comes from
uncertainty in $\eta$, the {\it slip factor} relating frequency and
momentum excursions in the storage ring \cite{psipaper}, \cite{mcginnis},
which is used to determine the $FWHM_{\sqrt{s}}$. That in
$B_{in}\Gamma_{out}$ is estimated by adding in quadrature the errors in
detector and luminosity-monitor acceptance and efficiency.  The
external-parameter error in $M_{\chi_{cJ}}$ comes from the uncertainty in
the $\psi'$ mass used in the absolute calibration of the beam energy
\cite{psipaper} as described above. For $B_{\bar p p}\Gamma_{J/\psi X }$
and $B_{\bar p p}\Gamma_{J/\psi \gamma}$, the uncertainties in the
parameters of the elastic scattering cross section \cite{elastic}, which
are the limiting errors in the estimate of luminosity, the parameters of
the final state angular distributions \cite{angdis}, and the branching
ratio $B_{J/\psi\rightarrow e^+ e^-}$, where we use $0.0593\pm 0.001$, are
added in quadrature.  The error contributions are summarized in Table
\ref{tab:errori}.

\begin {table*}[htb]
\begin{center}
\begin {tabular}{|ccc|}
\hline\hline
 & & \\
 & $\bar p p \rightarrow \chi_{c1} \rightarrow J/\psi + X $ &
 $\bar p p \rightarrow \chi_{c1} \rightarrow J/\psi + \gamma$ \\
 
 & & \\ 

 \hline

$M_{\chi_{c1}}$ [MeV/c$^2$] & $ 3510.719\pm 0.051 \pm 0.019_{ext}$ & 
$3510.713 \pm 0.051 \pm 0.019_{ext}$ \\
 
$\Gamma_{\chi_{c1}}$~[MeV] & $0.876 \pm 0.045 \pm 0.026_{aux}$ & $0.881
\pm 0.052 \pm 0.026_{aux}$ \\

B$_{\bar p p}\times \Gamma_{out}$~[eV] & $21.5\pm 0.5 \pm 0.6_{aux} \pm
0.6_{ext}$ & $19.8\pm 0.5\pm 0.6_{aux}\pm 0.6_{ext}$ \\

$\sigma_b$[pb] &$10.5\pm1.0$ & $2.5\pm 0.4$\\

\hline\hline

& & \\

 & $\bar p p \rightarrow \chi_{c2} \rightarrow J/\psi + X$  & 
$\bar p p \rightarrow \chi_{c2} \rightarrow J/\psi + \gamma$   \\
& & \\

\hline

$M_{\chi_{c2}}$ [MeV/c$^2$] & $3556.173 \pm 0.123 \pm 0.020_{ext}$ & 
$3556.168 \pm 0.114 \pm 0.020_{ext}$ \\

$\Gamma_{\chi_{c2}}$~[MeV] & $1.915 \pm 0.188 \pm 0.013_{aux}$ & $1.953\pm 
0.187 \pm 0.013_{aux}$ \\

B$_{\bar p p}\times \Gamma_{out}$~[eV] & $27.0 \pm 1.5 \pm 0.8_{aux} \pm 
0.7_{ext}$ & $26.8 \pm 1.7 \pm 0.8_{aux} \pm 0.8_{ext}$\\

$\sigma_b$[pb] & $12.0 \pm 1.3$ & $2.6\pm0.7$\\

\hline\hline
\end{tabular}
\vspace{+0.5cm}

\caption{E835 results for $\chi_{c1}$ and $\chi_{c2}$ resonance parameters
for the J/$\psi + X $ and J/$\psi + \gamma $ final states. The first
errors are statistical; the second and third are auxiliary-variable and
external-parameter systematic errors given in Table \ref{tab:errori}.}

\protect{\label{tab:chi12}}
\end{center}
\end{table*}

\begin {table}[htb]
\begin{center} 
\begin {tabular}{|ccccc|} 
\hline\hline 
Resonance Parameters & & Auxiliary Variables & & \\ 
\hline    & $\eta $    & Effic.    & Lumin.  & Total\\
$\Gamma_{\chi_{c1}}$ &  26 keV &  -             &   -  & 26 keV \\
$\Gamma_{\chi_{c2}}$ &  13 keV &  -             &   -  & 13 keV \\
$B_{\bar p p}\Gamma(\chi_{cJ}\rightarrow J/\psi\gamma)$  
 &  - & 3.0$\%$ & 0.6$\%$ &  3.1$\%$ \\
 $B_{\bar p p}\Gamma(\chi_{cJ}\rightarrow J/\psi $ X ) 
 & -  & 2.8$\%$      & 0.6$\%$ & 2.9$\%$ \\
\end {tabular}
\begin{tabular} {|cccccc|}
\hline\hline
Resonance Parameters & & External Parameters & & &\\
\hline     & M$_{\psi'}$   & a$_2$,B$_0$    & $\sigma_{tot},b, \rho $  & 
$B(J/\psi \rightarrow e^+ e^-)$ & Total\\
$M_{\chi_{c1}} $ & 19 keV/c$^2$ &  -             &   -  & & 19 keV/c$^2$ \\
$M_{\chi_{c2}}$ & 20 keV/c$^2$ &  -             &   -  & & 20 keV/c$^2$ \\
$B_{\bar p p}\Gamma(\chi_{c1}\rightarrow J/\psi\gamma)$ 
  & -  &  -   & 2.1$\%$ & 1.7$\%$ & 2.7$\%$ \\
$B_{\bar p p}\Gamma(\chi_{c2}\rightarrow J/\psi\gamma)$   
  & -  & 1.3$\%$ & 2.1$\%$ & 1.7$\%$ & 3.0$\%$\\
 $B_{\bar p p}\Gamma(\chi_{c1}\rightarrow J/\psi $ X) 
 & -  &  -  & 2.1$\%$ & 1.7$\%$ &  2.7$\%$ \\
$B_{\bar p p}\Gamma(\chi_{c2}\rightarrow J/\psi $ X)  
 & -  &  0.5$\%$ & 2.1$\%$ & 1.7$\%$ & 2.7$\%$ \\ 
\hline\hline
\end{tabular}
\vspace{+0.5cm} 

\caption{Uncertainties due to errors in auxiliary variables and external
parameters.  In the top panel we refer to variables measured during data
taking, while the bottom panel refers to parameters measured by other
experiments:  $\eta$ is the {\it slip factor} relating frequency and
momentum excursions in the storage ring; $a_2$ and $B_0$ characterize the
$\chi_c$ decay angular distributions \cite{angdis}; $\sigma_{tot}$, $b$,
and $\rho$ parametrize the $\bar{p}p$ elastic cross section
\cite{elastic}. Within each category, the contributions are added in
quadrature.}

\protect{\label{tab:errori}}
\end{center}
\end{table}

\begin{figure*}[hbtp]
\vspace*{+0.5cm}
\begin{center}
\begin{tabular}{c}
\includegraphics[scale=0.90 ] {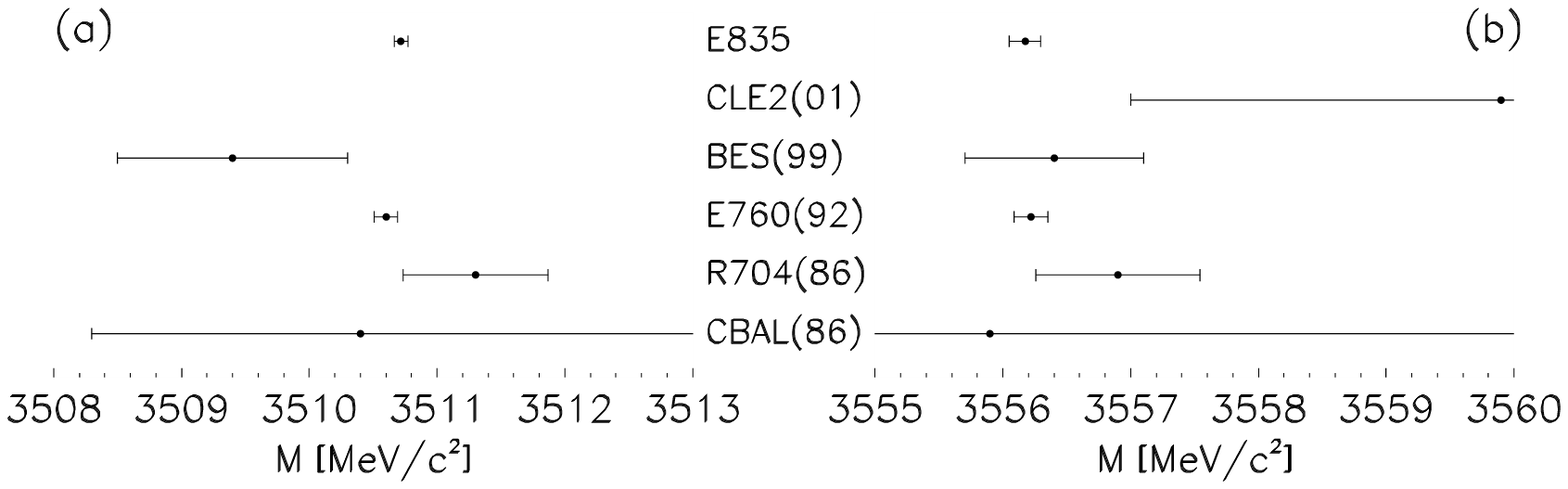}
\end{tabular}
\vspace{-0.4cm} 

\caption{$\chi_{c1}$ (a) and $\chi_{c2}$ (b) masses from this and other
%experiments. The full-length error bars include systematics.}
experiments. The error bars represent the in-quadrature sums of 
statistical and systematic errors.}

\label{fig:paragone}
\end{center}
\end{figure*}

In Fig. \ref{fig:paragone} a) and b) we compare the results of our mass
measurements for $\chi_{c1}$ and $\chi_{c2}$ to the values obtained in
other experiments in the last twenty years.  The comparison clearly shows
the advantage of using this technique.  For the width measurements, the
only results of precision comparable to those of E835 were obtained by our
predecessor experiment, E760. These are listed in Table \ref{tab:parag}.
  
\section {Discussion}

Our experimental study of charmonium is based on three data taking periods
in the years 1990-1991 (E760)\cite{chi12}, 1996-1997 (Run 1 of E835) and
2000 (Run 2 of E835).  Table \ref{tab:parag}, where we compare E835 with
E760, shows good agreement for all measured parameters.  In the nine year
gap between the first and last data-taking periods, major modifications of
the Accumulator lattice and of the beam diagnostic system took place and
elements of the target-detector complex were substituted or upgraded.  We
consider the consistency between the two experiments evidence of our
understanding of the related systematic uncertainties. \\

In Table \ref{tab:parag}, the values measured by E760 are adjusted as
follows: $M_{\chi_{cJ}}$ are adjusted to reflect the KEDR high-precision
$\psi'$ mass measurement referred to above; $M_{\chi_{cJ}}$,
$\Gamma_{\chi_{cJ}}$ and $B_{\bar p p}\times \Gamma(\chi_{cJ}\rightarrow
J/\psi +\gamma)$ are adjusted by the small shifts induced by fitting to
the Breit-Wigner cross section corrected for radiative effects (see
Appendix A and Table \ref{tab:radcorr}); $B_{\bar p p}\times
\Gamma(\chi_{cJ}\rightarrow J/\psi +\gamma)$ are corrected for a $6.5\%$
underestimate of the luminosity ({\it Ldt}), which we discovered after
publication of the results. The orbit uncertainty from the BPM system is
included in the statistical error and the systematic error in
$M_{\chi_{cJ}}$ comes entirely from the $\psi'$ mass uncertainty; the
systematic error in $\Gamma_{\chi_{cJ}}$ comes entirely from the
uncertainty in $\eta$, and that in $B_{\bar p p}\times
\Gamma(\chi_{cJ}\rightarrow J/\psi +\gamma)$ has contributions from
auxilliary variables and external parameters. For completeness, we include
the results from our recent measurement of the resonance parameters of the
$\chi_{c0}$ state \cite{chi0}. These values are adjusted for the KEDR
$\psi'$ mass and radiative effects.

Using the E835 values, we derive (Table \ref{tab:derivati}) the fine
structure splittings $\Delta M_{21} = M_{\chi_{c2}}-M_{\chi_{c1}}$ and
$\Delta M_{10} = M_{\chi_{c1}}-M_{\chi_{c0}}$, the ratio $\rho$ =
$\frac{\Delta M_{21}}{\Delta M_{10}}$, and the $\chi_{cJ}$ center of
gravity, $ M_{c.o.g.} = \frac{M_{\chi_{c0}} + 3M_{\chi_{c1}} +
5M_{\chi_{c2}}}{9}$. The uncertainties contain the statistical and 
systematic errors added in quadrature, accounting for systematic errors 
in common.\\

\begin {table*}[htb]
\begin{center}
\begin {tabular}{|ccc|}

\hline\hline
&{\large{$\chi_{c0}$}}&\\
\hline
PARAMETER         & E835       &     \\
$M_{\chi_{c0}}$ [MeV/c$^2$]   & $3415.5 \pm 0.4 \pm 0.4$  &  \\
$\Gamma_{\chi_{c0}}$~[MeV] & $9.7 \pm 1.0 $ &  \\
$B_{\bar p p}\times  \Gamma(\chi_{c0}\rightarrow J/\psi +\gamma)$~[eV] & 
$28.0 \pm 1.9 \pm 1.3$ & \\

\hline\hline
&{\large{$\chi_{c1}$}}&\\
\hline
PARAMETER         & E835       & E760      \\
$M_{\chi_{c1}}$ [MeV/c$^2$]   & $3510.719 \pm 0.051 \pm 0.019_{ext}$        
&  $ 3510.60 \pm 0.087 \pm 0.019$\\
$\Gamma_{\chi_{c1}}$~[MeV] & $0.876 \pm 0.045 \pm 0.026_{aux}$ & 
$0.87 \pm 0.11 \pm 0.08$\\
$B_{\bar p p}\times  \Gamma(\chi_{c1}\rightarrow J/\psi +\gamma)$~[eV]    
& $21.5 \pm 0.5 \pm 0.6_{aux} \pm 0.6_{ext}$ & $21.4 \pm 1.5 \pm 2.2$\\

\hline\hline
& {\large{$\chi_{c2}$}} & \\
\hline
PARAMETER          & E835           & E760                    \\
$M_{\chi_{c2}}$ [MeV/c$^2$] & $3556.173\pm 0.123 \pm 0.020_{ext}$ & 
$3556.22 \pm 0.131 \pm 0.020$ \\
$\Gamma_{\chi_{c2}}$~[MeV] & $1.915 \pm 0.188 \pm 0.013_{aux}$ 
& $1.96\pm 0.17 \pm 0.07$ \\
$B_{\bar p p}\times 
\Gamma(\chi_{c2}\rightarrow J/\psi +\gamma)$~[eV] &
 $27.0 \pm 1.5 \pm 0.8_{aux} \pm 0.7_{ext} $ 
&  $27.7 \pm 1.5 \pm 2.0$\\

\hline\hline
\end{tabular}
\vspace{+0.5cm}

\caption{Comparison of the values of the $\chi_{c1}$ and $\chi_{c2}$
parameters measured in E835 with those measured by E760. The E835
$\chi_{c0}$ values are included for completeness\cite{chi0}. The mass and
width values come from the $\bar p p \rightarrow J/\psi + X$ fits. The
E760 and E835 $\chi_{c0}$ values are referred to the KEDR $\psi'$ mass and
radiative corrections are made. The E760 values are corrected for the
luminosity error described in the text.}

\protect{\label{tab:parag}}
\end{center}
\end{table*}

\begin {table}[htb]
\begin{center}
\begin {tabular}{|cc|}
\hline\hline
$\Delta M_{21} = M_{\chi_{c2}}-M_{\chi_{c1}}$ [MeV/c$^2$] & $45.45 \pm 0.15$ \\
$\Delta M_{10} = M_{\chi_{c1}}-M_{\chi_{c0}}$ [MeV/c$^2$] & $95.2 \pm 0.6$ \\
$\rho$ = $\frac{\Delta M_{21}}{\Delta M_{10}}$ & $0.477 \pm 0.002$ \\
$M_{c.o.g.} = \frac{ M_{\chi_{c0}} + 3M_{\chi_{c1}} + 
 5M_{\chi_{c2}} }{9}$ [MeV/c$^2$]& $3525.39 \pm 0.10$\\
\hline\hline
\end{tabular}
\vspace{+0.5cm}

\caption{Fine-structure splittings ($\Delta M_{ik}$), ratio of
fine-structure splittings ($\rho$) and $^3 P_J$ states center-of-gravity
($M_{c.o.g.})$, as derived from mass values measured by E835.}

\label{tab:derivati}
\end{center}
\end{table}

In the Breit-Fermi theory, the $\chi_{cJ} $ and $h_c$ masses can be 
written as:

\begin{equation} 
M_J = M_0 + \langle h_{LS}\rangle\times\langle\vec{L}.\vec{S}\rangle_J + 
\langle h_{T}\rangle\times\langle S_{12}\rangle_J
\end{equation}
 where the three terms are, respectively, the expectation values of the
spin-independent, spin-orbit and tensor components of the $\bar{c} c$
Hamiltonian \cite{Lucha}. If we assume the $\chi_c$ to be pure
($\bar{c}c$) states and their spatial wavefunctions to be identical, M$_0$
is the same for the three states and the mass splittings yield the values
of $\langle h_{LS}\rangle $ and $\langle h_{T}\rangle $. As $$\langle
\vec{L}.\vec{S} \rangle_J = -2,-1, 1 $$ and $$\langle S_{12}\rangle_J =
12(\frac{(\vec{S}_1.\vec{r})(\vec{S}_2.\vec{r})}{r^2} -
\frac{\vec{S}_1.\vec{S}_2}{3}) = -4, 2, -\frac{2}{5}$$ for J = 0,1,2
respectively, we obtain the spin-orbit contribution:

\begin{equation}
\langle h_{LS}\rangle =  
\frac{2\Delta M_{10}+5\Delta M_{21}}{12} =  34.80 \pm0.09~{\rm MeV/c}^2 
\label{eq,aaat the $}
\end{equation}
 and the tensor contribution:

\begin{equation}
\langle h_{T}\rangle =  \frac{10\Delta M_{10}-5\Delta M_{21}}{72} = 
10.06 \pm 0.06~{\rm MeV/c}^2.  
\label{eq,kk}
\end{equation}

Assuming (as above) that the spatial wave functions of the $\chi_{cJ}$
states are identical, the partial widths for the E1 transitions
$\chi_{cJ}\rightarrow J/\psi \gamma$ are expected to scale as
$E_\gamma^3$. Our recent measurement of $\Gamma_{\chi_{c0}}$ \cite{chi0}
and improved measurements of $B(\chi_{c0}\rightarrow J/\psi\gamma)$
\cite{pdg04} allow us to test this prediction for all three $\chi_J$. We 
have performed a fit to the $\psi'$ and $\chi_{cJ}$ branching 
ratios analogous to that described in Reference \cite{pdg04} to obtain 
the radiative widths, which are given in Table \ref{tab:radwidth}. These 
are in agreement with $E_\gamma^3$ scaling.

\begin {table}[htb]
\begin {tabular}{|cccc|}
\hline\hline
   &  $\chi_{c0} $ & $\chi_{c1} $ & $\chi_{c2} $ \\
\hline
$E_{\gamma}$ [MeV] & 304 & 389 & 430  \\
$\Gamma(\chi_{cJ}\rightarrow J/\psi \gamma)$ [keV] & 
%$114 \pm 18$ & $278 \pm 33$ & $388 \pm 50$ \\
$119 \pm 16$ & $280 \pm 32$ & $416 \pm 34$ \\
$\Gamma(\chi_{cJ}\rightarrow J/\psi \gamma)/E_{\gamma}^3$ 
[$10^{-9}$MeV$^{-2}$]  & 
%$ 4.09\pm 0.64$ & $ 4.71\pm 0.56 $ & $ 4.89 \pm 0.64$ \\
$ 4.12\pm 0.57$ & $ 4.74\pm 0.54 $ & $ 5.24 \pm 0.43$ \\

\hline\hline
\end{tabular}
\vspace{+0.5cm}
\caption{Partial widths for the E1 transitions $\chi_{cJ}\rightarrow 
J/\psi \gamma$ showing agreement with $E_\gamma^3$ scaling. These values 
are obtained from a global fit to the data tabulated in Reference 
\cite{pdg04} and our results.} \protect{\label{tab:radwidth}}
\end{table}

\section{Summary}

Fermilab experiment E835 and our earlier experiment, E760, have measured
the resonance parameters and $B(\bar{c}c\rightarrow \bar{p}p) \times
\Gamma (\bar{c}c \rightarrow $ final state) for charmonium states formed
in antiproton-proton annihilations.  We have directly determined the
masses and widths of these states with unprecendented precision in
extremely low-background conditions.

In this paper we compile the resonance parameters of the $^3P_J$ states.  
We report new measurements of $\chi_{c1}$ and $\chi_{c2}$ detected
through the decay channels $J/\psi + anything $ and $ J/\psi + \gamma$,
and find excellent agreement between these results and those obtained
by experiment E760. From the mass measurements we derive the
fine-structure splittings between $\chi_{c0}$, $\chi_{c1}$, and
$\chi_{c2}$ with a precision of a fraction of a percent. We find that the 
radiative widths for the $E1$ transitions $\chi_{cJ}\rightarrow 
J/\psi\gamma$ scale as $E_\gamma^3$ as expected. 

\section{Acknowledgments}

We gratefully acknowledge the support of the Fermilab staff and
technicians and especially the Antiproton Source Department of the
Accelerator Division and the On-Line Department of the Computing Division.  
We wish to thank also the INFN and university technicians and engineers
from Ferrara, Genova, Torino and Northwestern for the valuable work done.
This research was supported by the Italian Istituto Nazionale di Fisica
Nucleare (INFN) and U.S. Department of Energy.\\

\section{Appendix A}

In the analysis of an excitation profile the Breit-Wigner resonance cross
section must be corrected to account for the radiation of the incoming
particle in the electromagnetic field of the target particle.  For a $\bar
p p $ initial state, D. C. Kennedy \cite{kennedy} has derived the
following expression for the corrected Breit-Wigner cross-section:

\begin{equation}
 \sigma_{BW}^{rad}(\beta,s)= \beta\int_0^{\sqrt{s}\over 2} 
{dk\over k}
({2k\over\sqrt{s}})^\beta \sigma_{BW}(s-2k\sqrt{s})
\end{equation}
 with 
\begin{equation} 
\beta = \frac{2\alpha}{\pi} \times[
\frac{s-2m_p^2}{\sqrt{s\times(s-4m_p^2)}} \times ln( \frac{s +
\sqrt{s\times(s-4m_p^2)}}{s -\sqrt{s\times(s-4m_p^2)}})- 1].
\end{equation} 
In our analysis we have convolved the corrected Breit-Wigner cross section
with the beam distribution at each point in the scan;  in this way we
properly account for the conditions of data taking.  As the radiated
photon energy falls between zero and half the total energy, and we wish to
correct $\Gamma_R$ at the one-percent level, which means
0.01/3500=3$\times$ 10$^{-6}$ of the total energy, particular care was
taken to avoid rounding errors in performing the numerical integration.
In Table \ref{tab:radcorr} we list, for the three $\chi_c $ states, the
change in the measured parameters resulting from the application of
radiative corrections. These changes are significantly smaller than the 
uncertainties in the measurements.

\begin {table}[htb]
\begin {tabular}{|cccc|}
\hline\hline
   &  $\chi_{c0} $ & $\chi_{c1} $ & $\chi_{c2} $ \\
\hline
$M_{\chi_{cJ}}$ [MeV/c$^2$] & -0.06 & - 0.01 & -0.02  \\
$\Gamma_{\chi_{cJ}}$ & -1.2 $\%$   & -1.1 $\%$ & -0.9 $\% $\\
$B_{\bar p p}\times\Gamma(\chi_{cJ}\rightarrow J/\psi +\gamma)$  
& +3.2 $\%$  &  +5.0 $\%$  & +4.5 $\% $   \\
\hline\hline
\end{tabular}
\vspace{+0.5cm}
\caption{Shifts in the values of the resonance parameters when radiative 
corrections are applied}
\protect{\label{tab:radcorr}}
\end{table}

%\end{linenumbers}        


\begin{thebibliography}{00}
\bibitem{charge}Our detector did not determine the sign of the particle
 charge; we assume the charges of the two electrons in these events to be
 opposite.
\end{thebibliography}
\end{document}